\documentclass[aps,prl,reprint,superscriptaddress]{revtex4-1}

\usepackage{xcolor}
\usepackage{tikz}

\usepackage{amssymb,amsmath}
\usepackage{graphicx}
\usepackage{upgreek}
\usepackage[english]{babel}
\usepackage[T1]{fontenc}
\usepackage[utf8]{inputenc}
\usepackage{graphicx}
\usepackage{bm}
\def\pmb#1{\setbox0=\hbox{#1}
\kern-.025em\copy0\kern-\wd0 \kern-.05em\copy0\kern-\wd0
\kern-.025em\raise.0433em\box0}

\newcommand{\C}{\boldsymbol{C}}
\newcommand{\W}{\boldsymbol{W}}

\begin{document}

\title{On static chiral Milton-Briane-Willis continuum mechanics}

\author{Muamer Kadic}
\address{Institut FEMTO-ST, UMR 6174, CNRS, Universit\'{e} de Bourgogne Franche-Comt\'{e}, 25000 Besan\c{c}on, France}
\address{Institute of Nanotechnology, Karlsruhe Institute of Technology (KIT), 76128 Karlsruhe, Germany}
\author{Andr\'e Diatta}
\address{Institute of Nanotechnology, Karlsruhe Institute of Technology (KIT), 76128 Karlsruhe, Germany}
\author{Tobias Frenzel}
\address{Institute of Applied Physics Karlsruhe Institute of Technology (KIT), 76128 Karlsruhe, Germany}
\author{Sebastien Guenneau}%
\address{Aix$-$Marseille Univ, CNRS, Centrale Marseille, Institut Fresnel, 13013 Marseille, France}
\author{Martin Wegener}
\address{Institute of Nanotechnology, Karlsruhe Institute of Technology (KIT), 76128 Karlsruhe, Germany}
\address{Institute of Applied Physics Karlsruhe Institute of Technology (KIT), 76128 Karlsruhe, Germany}
\email{muamer.kadic@gmail.com}



\begin{abstract}
Recent static experiments on twist effects in chiral three-dimensional mechanical metamaterials have been discussed in the context of micropolar Eringen continuum mechanics, which is a generalization of Cauchy elasticity. For cubic symmetry, Eringen elasticity comprises nine additional parameters with respect to Cauchy elasticity, of which three directly influence chiral effects. Here, we discuss the behavior of the static case of an alternative generalization of Cauchy elasticity, the Milton-Briane-Willis equations. We show that in the homogeneous static cubic case only one additional parameter with respect to Cauchy elasticity results, which directly influences chiral effects. We show that the Milton-Briane-Willis equations qualitatively describe the experimentally observed chiral twist effects, too. We connect the behavior to a characteristic length scale.
\end{abstract}

\maketitle	
\section{1. Introduction}
In one dimension, the scalar spring constant in Hooke's law connects forces and displacements. As a generalization towards three-dimensional continuum mechanics \cite{Banerjee2011}, the rank-4 Cauchy elasticity tensor, ${\bf \C}$, connects the rank-2 stress tensor, $\boldsymbol{\sigma}$, and the rank-2 strain tensor, $\boldsymbol{\epsilon}$. In general, Cauchy elasticity comprises up to 21 independent nonzero parameters describing possible linear deformations of elastic bodies in three dimensions \cite{Sommerfeld1950,Authier2003,Miltonbook,Miltonbook2,Laude2015}. For cubic crystals, which are characterized by four three-fold rotational axes, only 3 parameters remain \cite{Authier2003}. 

However, Cauchy elasticity essentially only grasps the displacements, $\boldsymbol{u}(\boldsymbol{r}),$ of infinitesimally small volume elements (of ``points'') within a fictitious continuum. Cauchy elasticity therefore misses certain degrees of freedom in artificial three-dimensional periodic microlattices or metamaterials, for which the unit cell has a finite extent rather than being approximately point-like such as atoms in an ordinary crystal of macroscopic size \cite{Frenzel2017,Rueger2018}. Such missed degrees of freedom have recently become particularly obvious in chiral three-dimensionally periodic mechanical metamaterial structures (see Fig.\,1(a)) for which Cauchy elasticity fails to describe any effect of chirality, whereas prominent twist effects have been observed experimentally in the static case \cite{Frenzel2017}. In contrast, micropolar Eringen elasticity (see Fig.\,1(b)) has been able to describe these experimental findings \cite{Frenzel2017} as well as others for achiral media \cite{Rueger2018}. Cosserat elasticity \cite{Eringen1974} can be seen as a special case of Eringen micropolar elasticity.

\begin{figure}[h!]
	\includegraphics[width=8.5cm,angle=0]{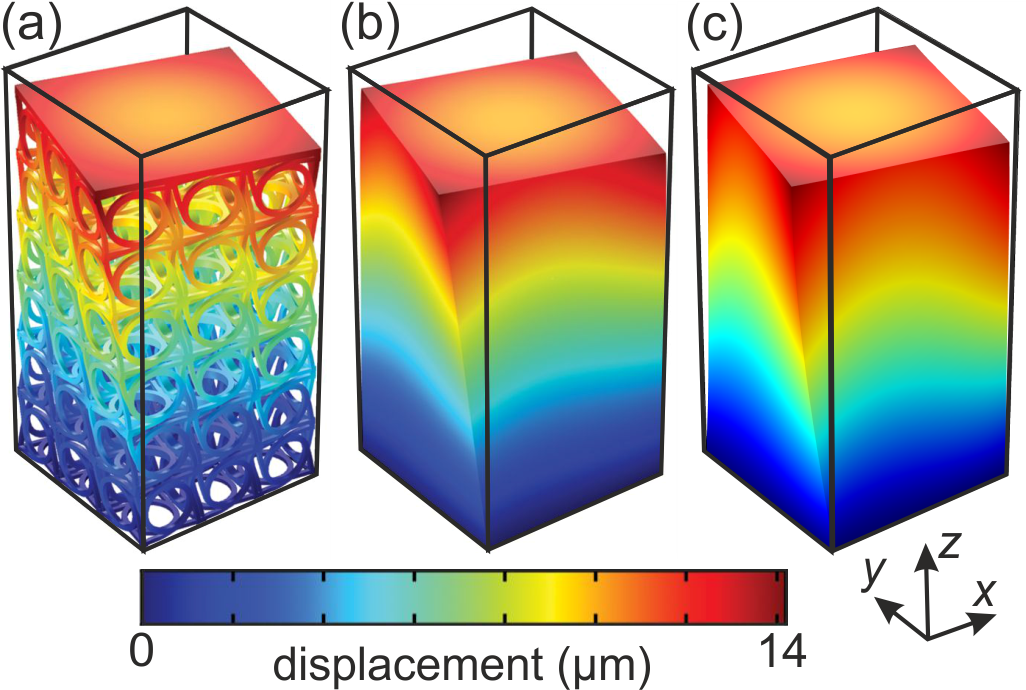}
	\caption{Cuboid beams with volume $L \times L \times 2L$, with side length $L=500\ \mathrm{\mu m}$, are subject to uniaxial loading along the negative $z$-direction. In Cauchy elasticity, the beam compresses and can expand or contract laterally, depending on its Poisson ratio (not depicted). However, a twist is forbidden in Cauchy elasticity, even if the underlying crystal symmetry would allow for a twist. (a) Finite-element calculation for a chiral metamaterial microstructure exhibiting a twist behavior \cite{Frenzel2017}. (b) Same as (a), but calculated using chiral micropolar Eringen elasticity \cite{Frenzel2017}. Panels (a) and (b) are taken with permission from \cite{Frenzel2017}. (c) Numerical calculations based on generalized Cauchy elasticity following Milton-Briane-Willis. For all three panels, the effects are calculated within the linear elastic regime and are magnified tenfold for clarity. The modulus of the displacement vector field is superimposed on a false-color scale. Parameters in (c) are: $C_{11}=32.8\; \rm MPa$, $C_{12}=-6.1\; \rm MPa$,  $C_{44}=19.4\; \rm MPa$ (in Voigt notation), and $\alpha =3 \;\rm GPa/m$.}
	\label{figure1}
\end{figure}

Cosserat elasticity, micropolar Eringen elasticity \cite{Bigoni2007}, micromorphic Eringen elasticity \cite{Eringen1974,toupin1964}, strain-gradient theories \cite{Gudmundson2004}, and yet more advanced approaches \cite{Seppecher2011} are not the only possible generalizations of Cauchy elasticity though. It is therefore interesting and relevant to ask which generalizations other than Eringen's can describe the effects of chirality observed in recent experiments \cite{Frenzel2017}. 

In this paper, in Section 2., we start with the static version of a generalization of Cauchy elasticity following Milton, Briane, and Willis \cite{Milton2006}. This generalization arises from demanding that the resulting equations are mathematically form invariant under coordinate transformations. Aiming at describing recent experiments \cite{Frenzel2017}, we focus on the case of three-dimensional homogeneous cubic crystals without centrosymmetry, in which case the terms beyond Cauchy elasticity can be parameterized by a single scalar parameter. In Section 3., we discuss numerical solutions. We find that the resulting behavior qualitatively describes the push-to-twist conversion effects observed in recent experiments (see Fig.\,1(c)) and that it can be connected to a characteristic length scale. We conclude in Section 4.

\section{2. Generalized Static Cauchy elasticity}

In the static case, all forces must balance. For simplicity, we omit external forces in all formulas throughout this paper; we implement them via boundary conditions. Hence, the divergence of the stress tensor $\boldsymbol{\sigma}=\C:\ \boldsymbol{\epsilon }$ is zero, ${\boldsymbol{\nabla }}\cdot \boldsymbol{\sigma }=\boldsymbol{0}$. The symbol ``:'' denotes a double contraction, the dot ``$\cdot$'' denotes a contraction between two tensors. Cauchy elasticity reduces to the compact equation
\begin{equation} \label{GrindEQ__1_} 
{\boldsymbol{\nabla}}\cdot \left(\C:\boldsymbol{\epsilon }\ \right)=\boldsymbol{0}, 
\end{equation} 
where $\C$ is the rank-4 elasticity tensor, with components $C_{ijkl}$ ($i,j,k,l=1,2,3$) in Cartesian coordinates and S.I. units of $\mathrm{Pa}$, and $\boldsymbol{\epsilon }$ is the dimensionless symmetric rank-2 strain tensor with components ${\epsilon }_{ij}={\epsilon}_{ji}$ \cite{Sommerfeld1950}. The strain tensor can be connected to the gradient of the displacement vector field $\boldsymbol{u}=\boldsymbol{u}(\boldsymbol{r})$ with components ${u}_i$ ($i=1,2,3$) via \cite{Sommerfeld1950}
\begin{equation} \label{GrindEQ__2_} 
\boldsymbol{\epsilon }=\frac{1}{2}\left({\boldsymbol{\nabla }}\boldsymbol{u}+{\left({\boldsymbol{\nabla }}\boldsymbol{u}\right)}^{\mathrm{T}}\right), 
\end{equation} 
where the superscript ``T'' refers to the transposed quantity. The Cauchy elasticity tensor obeys the minor symmetries (${\mathop{C}}_{ijkl}=C_{jikl}=C_{ijlk}$) and the major symmetries ($C_{ijkl}=C_{klij}$) \cite{Sommerfeld1950}. As a result, the strain tensor $\boldsymbol{\epsilon}$ in \eqref{GrindEQ__1_} can equivalently be replaced by the gradient of the displacement vector $\boldsymbol{\nabla}\boldsymbol{u}$, i.e.,
\begin{equation}\label{new}
{\boldsymbol{\nabla}}\cdot \left(\C:\boldsymbol{\nabla}\boldsymbol{u}\ \right)=\boldsymbol{0}\,.
\end{equation}

Cauchy elasticity does not describe effects of chirality at all \cite{Eringen1974}. This fact can immediately be seen by recalling that all even-rank tensors (such as the rank-2 stress tensor, the rank-4 elasticity stress tensor, and the rank-2 strain tensor) are invariant under space inversion operations, $\boldsymbol{r}\to -\boldsymbol{r}$ \cite{Eringen1974}. Thus, \eqref{new} does not change under space inversion, which brings one from a left-handed to a right-handed medium (or vice versa). 

A space inversion is a special example of a general spatial coordinate transformation $\boldsymbol{r}\to {\boldsymbol{r}'}\left(\boldsymbol{r}\right)$. It has first been pointed out by Milton, Briane, and Willis \cite{Milton2006} in 2006 that Cauchy elasticity is not form invariant under general curvilinear coordinate transformations (also see \cite{Brun2009,Norris2011,Diatta2014}). This finding is disturbing in view of the fact that most other equations in physics such as, for example, the quantum mechanical Schr\"{o}dinger equation, the wave equation of longitudinal pressure waves in gases/fluids, the time-dependent Maxwell equations, Fourier's time-dependent heat conduction equation, Fick's stationary diffusion equation, the stationary electric conduction equation, and the stationary laminar fluid-flow equation are form invariant under coordinate transformations \cite{Dollin,Leonhardt2006,Pendry2006,Milton2006,Cummer2007,Brun2009,Norris2011,Kadic2012b,Guenneau2012,Guenneau2013,McCall2018}. Milton, Briane, and Willis arrived at a generalization of dynamic Cauchy elasticity that is form invariant under coordinate transformations \cite{Milton2006}. In the static limit, i.e., for angular frequency $\omega=0$ and finite static mass density $\rho$, their equation (2.4) reduces to
\begin{equation} \label{GrindEQ__3_} 
{\boldsymbol{\nabla }}\cdot \left(\C: {\boldsymbol{\nabla}}\boldsymbol{u} +\boldsymbol{S}\cdot \boldsymbol{u}\right)-\boldsymbol{D}: {\boldsymbol{\nabla }}\boldsymbol{u}=\boldsymbol{0}, 
\end{equation} 
where $\C=\C(\boldsymbol{r})$ and the two additional rank-3 tensors $\boldsymbol{S}=\boldsymbol{S}(\boldsymbol{r})$ and $\boldsymbol{D}=\boldsymbol{D}(\boldsymbol{r})$ generally depend on the spatial position $\boldsymbol{r}$. 

In the case of a \textit{homogeneous material or homogenized structure} with ${\boldsymbol{\nabla }} S_{ijk}={\boldsymbol{\nabla }} D_{ijk}=\boldsymbol{0}$, which is the focus of interest in this paper, equation \eqref{GrindEQ__3_} reduces to 
\begin{equation} \label{GrindEQ__18_} 
\boldsymbol{\nabla}\cdot \left(\C: {\boldsymbol{\nabla}}\boldsymbol{u}\ +\W\cdot \boldsymbol{u}\right)=\boldsymbol{0}, 
\end{equation} 
with the rank-3 tensor $\W$ defined by
\begin{equation} \label{GrindEQ__19_} 
\W=\boldsymbol{S}-\boldsymbol{D}^{\mathrm{T}},           
\end{equation} 
where the components of the ``transposed'' tensor are given by 
\begin{equation} 
D^{\mathrm{T}}_{ijk}=D_{jik}.
\end{equation}

Broken centrosymmetry is a necessary requirement for chirality \cite{Eringen1974,Authier2003,Frenzel2017}. If we nevertheless consider an {\it isotropic medium or a cubic crystal with centrosymmetry}, it follows that $\W\equiv \boldsymbol{0}$, just like for any homogeneous rank-3 tensor \cite{Authier2003}. $\W\equiv \boldsymbol{0}$ also holds true for an {\it isotropic medium without centrosymmetry}.

For a \textit{cubic crystal without centrosymmetry}, we find that the tensor $\W$, such as any rank-3 tensor \cite{Authier2003}, reduces to the rank-3 Levi-Civita tensor
${\boldsymbol{\varepsilon}}$  (with components ${{\varepsilon}}_{123}={\mathrm{\varepsilon }}_{231}={\mathrm{\varepsilon }}_{312}=-{\mathrm{\varepsilon }}_{132}=-{\mathrm{\varepsilon }}_{213}=-{\mathrm{\varepsilon }}_{321}=1$, all other components are zero) times a scalar factor $\alpha $, i.e.,
\begin{equation} \label{GrindEQ__21_} 
\W=\alpha \, \boldsymbol{\varepsilon}. 
\end{equation} 
$\alpha $ has S.I. units of $\mathrm{Pa/m}$. 
This allows us to rewrite Milton-Briane-Willis elasticity \eqref{GrindEQ__18_} to 
\begin{equation} \label{GrindEQ__22_} 
\boldsymbol{\nabla} \cdot \left(\C:\ \boldsymbol{\nabla} \boldsymbol{u}\right)-\alpha \ \boldsymbol{\nabla}\times \boldsymbol{u}=\boldsymbol{0}. 
\end{equation} 

It is instructive to investigate the behavior of \eqref{GrindEQ__22_} under a space inversion operation, $\boldsymbol{r}\to -\boldsymbol{r}$. As argued below \eqref{new}, $\C:\boldsymbol{\nabla} \boldsymbol{u}=\C:\boldsymbol{\epsilon}$ does not change sign, but the $\boldsymbol{\nabla} $ in front does. In the second term in \eqref{GrindEQ__22_}, both $\boldsymbol{\nabla}$ and $\boldsymbol{u}$ do change sign, hence the vector product does not change sign. Therefore, the relative sign of the first and second term in \eqref{GrindEQ__22_} changes when performing a space inversion. Thus, \eqref{GrindEQ__22_} is different for a left- and a right-handed medium, respectively. This behavior is a necessary condition for a continuum formulation to be able to describe the effects of chirality in mechanics. Clearly, if the single parameter beyond Cauchy elasticity is zero, $\alpha=0$, equation \eqref{GrindEQ__22_} reduces to Cauchy elasticity \eqref{new}. As usual \cite{Authier2003,Banerjee2011}, for cubic symmetry (with or without a center of inversion), the Cauchy elasticity tensor $\C$ contains $3$ independent nonzero scalar parameters \cite{Authier2003}. 

We will use \eqref{GrindEQ__22_} for the numerical calculations presented in Section 3. To be unambiguous and clear for experimentalists, we therefore write equation \eqref{GrindEQ__22_} out into its three components and explicitly write out all involved sums, leading to 
\begin{eqnarray}
\sum^3_{i,k,l=1}{\left(\frac{\partial }{\partial x_i}\left(C_{i1kl}\frac{\partial u_l}{\partial x_k}\right)\right)}-\alpha \left(\frac{\partial u_3}{\partial x_2}-\frac{\partial u_2}{\partial x_3}\right)=0,\,\,\,\,\,\,\,\,\, \\
\sum^3_{i,k,l=1}{\left(\frac{\partial }{\partial x_i}\left(C_{i2kl}\frac{\partial u_l}{\partial x_k}\right)\right)}-\alpha \left(\frac{\partial u_1}{\partial x_3}-\frac{\partial u_3}{\partial x_1}\right)=0,\,\,\,\,\,\,\,\,\,
\end{eqnarray}
and
\begin{eqnarray}
\sum^3_{i,k,l=1}{\left(\frac{\partial }{\partial x_i}\left(C_{i3kl}\frac{\partial u_l}{\partial x_k}\right)\right)}-\alpha \left(\frac{\partial u_2}{\partial x_1}-\frac{\partial u_1}{\partial x_2}\right)=0.\,\,\,\,\,\,\,\,\, 
\end{eqnarray}

\section{3. Numerical calculations}

In what follows, we illustrate the generalized static homogeneous equation \eqref{GrindEQ__22_} for a cubic three-dimensional chiral medium by example numerical calculations. To allow for a direct comparison with micropolar Eringen elasticity, we reproduce previous continuum results \cite{Frenzel2017} and results of metamaterial microstructure calculations \cite{Frenzel2017} at selected points for convenience of the reader. To ease this comparison, we also choose similar parameters as much as possible.

\subsection{3.1 Numerical approach}

In our numerical calculations, we consider cuboid shaped samples with volume $L\times L\times 2L$. We apply uniaxial loading by a rigid stamp along the negative $z$-direction with sliding boundary conditions at the top, i.e., at the top surface we have the displacement $\boldsymbol{u}={\left(0,0,u_z\right)}^{\mathrm{T}}$ with $u_z \neq 0$. The axial strain results from $\epsilon=\epsilon_{33} =-u_z/L$. The uniaxial pressure, $P$, exerted at the top is given by $P={\boldsymbol{n}}\;\mathrm{\cdot}\;\mathrm{(}\C \;\mathrm{:}\;\boldsymbol{\nabla} \boldsymbol{u}-\alpha \ \boldsymbol{\nabla} \times \boldsymbol{u}  \mathrm{)}$, where ${\boldsymbol{n}}$ is the normal vector pointing into the negative $z$-direction. On the four sides we use open boundary conditions, i.e., ${\boldsymbol{n}}\;\mathrm{\cdot }\left(\C \; \mathrm{:} \;\boldsymbol{\nabla}\boldsymbol{u}-\alpha \ \boldsymbol{\nabla} \times \boldsymbol{u}  \right)\mathrm{=0}$, with the respective normal vectors ${\boldsymbol{n}}$ of the four side facets. On the bottom of the cuboid, we use fixed boundary conditions with $\boldsymbol{u}={\left(0,0,0\right)}^{\mathrm{T}}$, describing that the sample cuboid is fixed to a substrate. We have used the same conceptual boundary conditions in our previous work on static Eringen elasticity \cite{Frenzel2017}.

We solve equation \eqref{GrindEQ__22_} by using a finite-element approach via the partial differential equation (PDE) module of the commercial software package COMSOL Multiphysics. Herein, the homogeneous sample cuboid is typically discretized into $10^4$ tetrahedra, corresponding to about $5 \times 10^4$ degrees of freedom. Finer discretization has led to negligible changes with respect to the results outlined in the following. 

In Section 3.2., we will discuss the behavior of the twist angle and the axial strain of the cuboid sample under uniaxial loading. The axial strain is defined as the $z$-component of the displacement vector at the top surface (which is the same for all positions ${\left(x,y,2L\right)}^{\mathrm{T}}$), divided by the sample length $2L$, i.e., by $u_z(x,y,2L)/(2L)$.

The twist angle is defined via the displacement of the equivalent four corners at the top of the sample cuboid in the $xy$-plane, which are at positions ${\left(\pm L/2,\pm L/2,2L\right)}^{\mathrm{T}}$ before loading, with respect to the sample center at  ${\left(0,0,2L\right)}^{\mathrm{T}}$. For pure twists without further deformations, this definition grasps the entire sample behavior. If deformations that are more complex occur in addition, the twist angle resulting from our definition should be seen as merely a parameter representing part of the overall behavior.

\subsection{3.2 Results and discussion}

Figure 1 shows the modulus of the displacement vector field (on a false-color scale) for uniaxial loading along the negative $z$-axis of a cuboid shaped sample with volume $L\times L\times 2L$. All results shown are within the linear elastic regime, i.e., for axial strains $<1 \% $. We choose $L=500\, \rm \mu m$ (see $N=1$ in \cite{Frenzel2017}), $C_{11}=C_{22}=C_{33}$ (in standard Voigt notation \cite{Eringen1974}), $C_{12}=C_{13}=C_{21}=C_{23}=C_{31}=C_{32}$, $C_{44}=C_{55}=C_{66}$ (all other elements of the elasticity tensor are zero), and $\alpha =3\,$GPa/m. 

The results of Milton-Briane-Willis generalized Cauchy elasticity in Fig.\,1(c) are compared with those of micropolar Eringen elasticity \cite{Frenzel2017} in panel (b) and finite-element metamaterial microstructure calculations \cite{Frenzel2017} in panel (a). For the details underlying panels (a) and (b), we refer the reader to the extensive discussion in \cite{Frenzel2017} and the corresponding supporting online material. Obviously, Milton-Briane-Willis generalized Cauchy elasticity, micropolar Eringen elasticity, and the finite-element microstructure calculations exhibit the same qualitative behavior. When replacing $\alpha \to -\alpha $, the direction of the twist changes from clockwise to counter-clockwise in Figs.\,1(b) and (c) (not depicted), corresponding to the behavior of the mirror image of the 3D microstructure shown in Fig.\,1(a).

In Figure 2(a), the parameters are the same as in Fig.\,1(c), except that we consider the three choices (a) $\alpha =3\,$GPa/m, (b) $\alpha \to 3.3\,\alpha $, and (c) $\alpha \to 33\,\alpha $. In Fig.\,2(b), the twist effect is simply larger than that in panel (a). In panel (c), however, unusual additional substructures appear in the displacement field.  

\begin{figure}[h!]
	\includegraphics[width=8.5cm,angle=0]{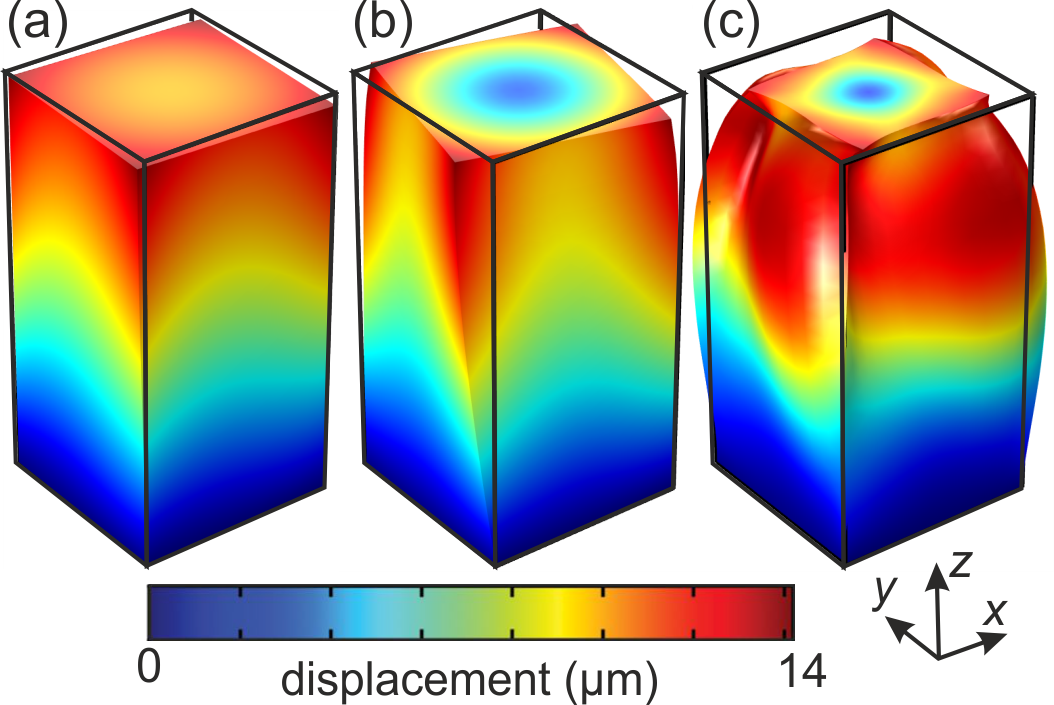}
	\caption{(a) Same as Fig.\,1(c) with parameter $\alpha =3\;$GPa/m and all other parameters fixed as in Fig.\,1(c). (b) $\alpha \to 3.3 \;\alpha $. (c) $\alpha \to 33 \;\alpha $. From such raw data, the twist angle at the top per axial strain can be deduced and plotted versus $\alpha $ (see Fig.\,3). All deformations are magnified twofold for clarity.}
	\label{figure2}
\end{figure}

Figure 3 emphasizes essentially the same aspect as Fig.\,2, however, we do not depict displacement fields of a sample but rather plot the calculated twist/strain (defined in Section 3.1.) versus the parameter $\alpha$ for fixed sample side length $L$. For small values of $\alpha$, the twist/strain increases monotoneously. However, for larger values of $\alpha$, we find an unusual non-monotoneous resonance-like behavior (compare \cite{milton07}), which is connected to the behavior shown in Fig.\,2(c). 

This behavior versus the parameter $\alpha$ for fixed sample side length $L$ is connected to the behavior versus $L$ for fixed $\alpha$. Following references \cite{Eringen1974,Frenzel2017}, the effects beyond static Cauchy elasticity should \textit{decrease} with increasing $L$. More specifically, the twist/strain should decrease proportional to the surface-to-volume ratio, i.e., decrease $\propto 1/L$. What one gets from \eqref{GrindEQ__22_} versus $L$ for fixed $\alpha$ is the polar opposite of this behavior. This can be seen as follows: If we replace the spatial components $x_i\rightarrow \zeta x_i$ in \eqref{GrindEQ__22_}, with some dimensionless scaling factor $\zeta$, the ratio of the second and first term in \eqref{GrindEQ__22_} increases by factor $\zeta$. This means that the effects beyond Cauchy elasticity would increase with increasing sample side length $L$ if $\alpha$ was constant. We conclude that $\alpha$ must \textit{not} be considered as a constant material parameter, but rather as an effective continuum-model parameter. 

To arrive at a meaningful material parameter, $\beta$, we make the ansatz
\begin{equation} \label{GrindEQ__26_} 
\alpha =\frac{\beta }{L^2}\ . 
\end{equation} 
The parameter $\beta $ has S.I. units of $\mathrm{Pa}\,\mathrm{m}$. Thus, the ratio 
\begin{equation} \label{GrindEQ__27_} 
l_{\mathrm{c}}=\frac{\beta}{C} 
\end{equation} 
has units of a length. Here, $C$ is a nonzero element of the elasticity tensor $\C$ or a combination of elements. As the twist effect mainly changes the shape of the specimen but not its volume, we choose the shear modulus $C=C_{44}=C_{2323}$. The length $l_{\mathrm{c}}$ is obviously zero in the Cauchy limit of $\alpha =\beta =0$. Therefore, it is tempting to interpret $l_{\mathrm{c}}$ as a characteristic length scale in the same spirit as characteristic length scales in micropolar Eringen elasticity \cite{Eringen1974}. There, one gets several different characteristic length scales, all of which are zero in the Cauchy limit. 

\begin{figure}[h!]
	\includegraphics[width=8.5cm,angle=0]{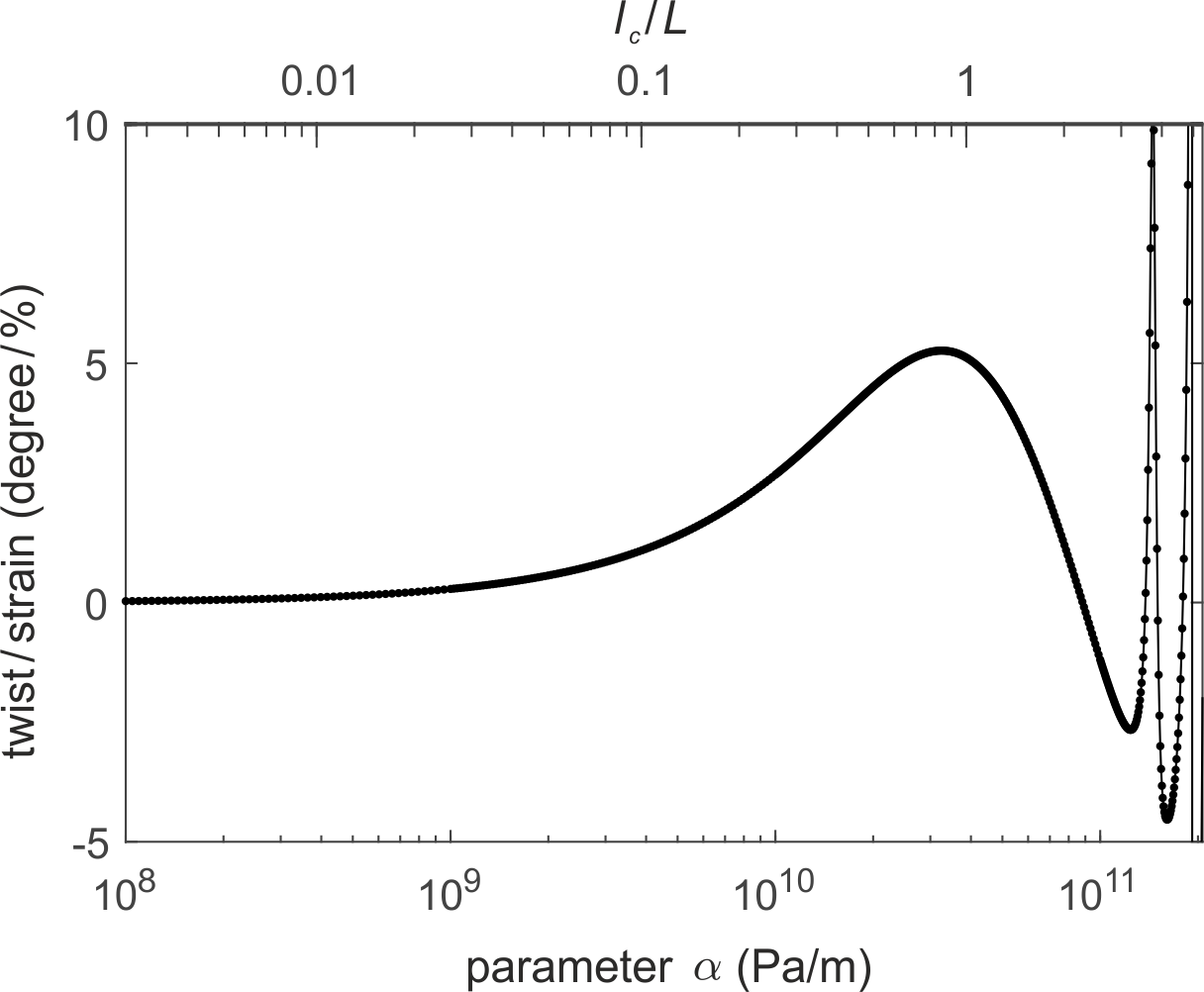}
	\caption{Twist/strain versus parameter $\alpha $ (lower horizontal scale) as deduced from raw data such as the ones shown in Fig.\,2. The dots are calculated, the curve is a guide to the eye. Parameters are as in Fig.\,1(c). The upper horizontal scale shows the characteristic length $l_{\mathrm{c}}$ normalized to the sample side length $L$, $\frac{l_{\mathrm{c}}}{L}=\frac{\beta /C_{44}}{L}=\alpha \frac{L}{C_{44}}$ according to \eqref{GrindEQ__28_}. Here, $C_{44}$ is an element of the elasticity tensor in Voigt notation, namely the shear modulus $C_{2323}$.}
	\label{figure3}
\end{figure}

To test this ansatz for the characteristic length scale, we depict as the upper horizontal scale in Fig.\,3 the normalized characteristic length, $l_{\mathrm{c}}/L$, which follows from 
\begin{equation} \label{GrindEQ__28_}
\frac{l_{\mathrm{c}}}{L}=\frac{\beta / C_{44}}{L}=\alpha \frac{L}{C_{44}}. 
\end{equation} 
We find that non-monotoneous behavior in Fig.\,3 occurs when $l_{\rm c}$ becomes comparable to or even exceeds the sample side length $L$. Likewise, the characteristic length $l_{\mathrm{c}}=1275 \, \rm \mu m$ in Fig.\,2(c) is also larger than the sample side length of $L=500\,\rm \mu m$, whereas $l_{\rm c}$ is smaller by factor $33$ and $10$, respectively, in panels (a) and (b).

\begin{figure}[h!]
	\includegraphics[width=8.5cm,angle=0]{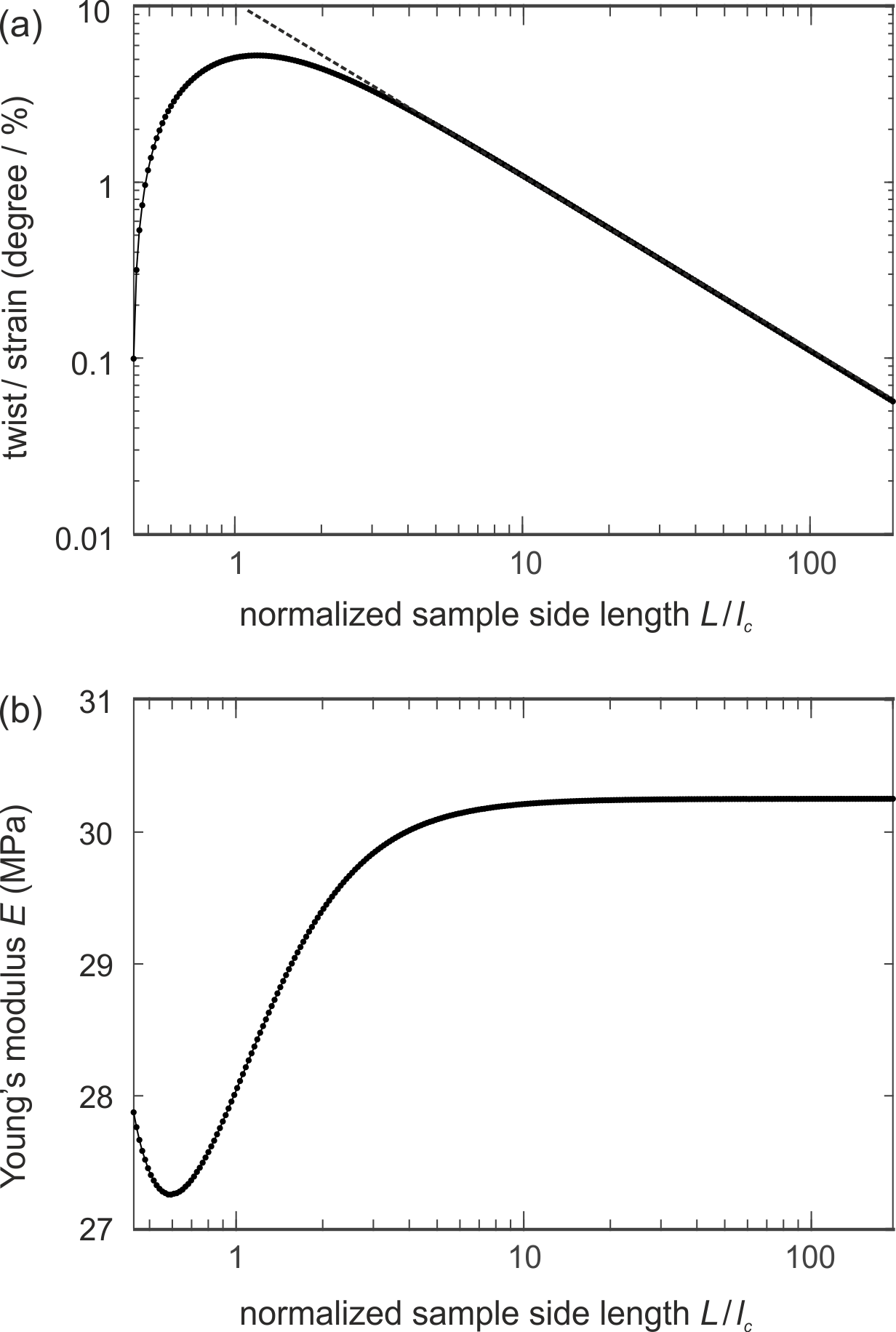}
	\caption{(a) Twist/strain versus sample side length $L$ normalized by the characteristic length $l_{\rm c}$ on a double logarithmic scale. The dependence on the parameter $\beta$ is implicitly contained in this normalization. The dashed straight line has a slope of $-1$ corresponding to the expected asymptotic scaling of the twist angle $\propto 1/L$ for fixed parameter $\beta$. (b) Same as in (a) but for the effective Young's modulus $E$ on a semi-logarithmic scale. All other parameters are as in Fig.\,1(c). The dots are calculated, the curves are guides to the eye. }
	\label{figure4}
\end{figure}

Finally, we study the dependence of the behavior on the sample side length $L$ for fixed parameter $\beta$ and fixed elements of the Cauchy elasticity tensor $\C$ (hence fixed $l_{\mathrm{c}}$) in Figure 4. In its panel (a), we plot the twist/strain versus sample side length $L$ on a double-logarithmic scale and in panel (b) the effective Young's modulus $E$ versus $L$ on a semi-logarithmic scale. The twist/strain in Fig.\,4(a) decreases inversely proportional to $L$ in the limit $L \gg l_{\rm c}$ (compare dashed straight line). The Young's modulus in panel (b) initially increases until it reaches a constant level for large values of $L$. Thereby, Cauchy elasticity, for which the twist is zero and the Young's modulus is independent on sample side length, is recovered in the large-sample limit of $L\to \infty $ in panels (a) and (b) of Fig.\,4 -- as it should. 

This overall behavior is qualitatively closely similar to the one which we have recently found in numerical calculations on chiral micropolar Eringen elasticity \cite{Eringen1974} as well as in our experiments on chiral three-dimensional metamaterials \cite{Frenzel2017}, all of which have been performed under static conditions. However, the quantitative agreement with experiments is worse for Milton-Briane-Willis elasticity than for Eringen elasticity. For example, the effective Young’s modulus $E$ versus sample side length $L$ increased by about a factor of ten in \cite{Frenzel2017} before it reached a constant value, whereas it only increases by about $10\%$ before it reaches a constant value in Fig.\,4(b). Moreover, the displacement field for the three-dimensional microstructure in Fig.\,1(a), which agrees quantitatively with Eringen elasticity in Fig.\,1(b), shows a somewhat more pronounced minimum in the middle of the sample top facet for Milton-Briane-Willis elasticity in Fig.\,1(c).

\section{4. Conclusion}

In conclusion, we have considered the static version of a generalized form of Cauchy elasticity, the Milton-Briane-Willis equations, for the case of three-dimensional homogeneous chiral non-centrosymmetric cubic media, which have been subject of recently published experimental and numerical work on mechanical metamaterials. We have found that this form of generalized static Cauchy elasticity grasps the quintessential qualitative features of recent experiments and of chiral micropolar Eringen elasticity, however, with just a single additional parameter. Under the same conditions, Eringen elasticity comprises $9$ additional parameters with respect to Cauchy elasticity, $3$ of which directly influence chiral effects.

Such non-uniqueness of effective-medium models is common for advanced continuum descriptions of materials, e.g., in electromagnetism and optics. It will be interesting to see in the future inasmuch Milton-Briane-Willis elasticity is able to describe more advanced static experiments or aspects of dynamic wave propagation in experiments on three-dimensional chiral mechanical metamaterials, and whether distinct qualitative differences with respect to micropolar Eringen elasticity arise or not. 

\section*{Acknowledgements}

We thank Graeme W. Milton (Utah, USA), Jensen Li (Birmingham, UK), and Christian Kern (KIT, Germany) for stimulating discussions. We acknowledge support by the German Federal Government, by the State of Baden-W\"{u}rttemberg, by KIT, and by the University Heidelberg through the Excellence Cluster ``3D Matter Made to Order (3DMM2O)'', by the Helmholtz program ``Science and Technology of Nanosystems (STN)'', and by KIT via the Virtual Materials Design project (VIRTMAT).
M.K. acknowledeges support by the EIPHI Graduate School (contract ''ANR-17-EURE-0002'') and the French Investissements d'Avenir program, project ISITE-BFC (contract ANR-15-IDEX-03).

%

%

\end{document}